# The start of things for Kamiokande: The Kamioka Nucleon Decay Experiment

Atsuto Suzuki

*Iwate Prefectural University, Sugo, Takizawa, Iwate, 020-0693, Japan*
E-mail: suzuki@iwate-pu.ac.jp

In 1980, detecting the evidence of the so-called Grand Unified Theory (GUT) became a major challenge for experimentalists. The GUT predicted that protons would decay, and M. Koshiba proposed the Kamioka Nucleon Decay Experiment (Kamiokande), which was originally conceived and designed for the detection of proton decay signals. Kamiokande was a 3000-t water Cherenkov detector located 1000 m underground in the Kamioka mine. The unique feature of this experiment was the use of 1000 photomultiplier tubes with the world's largest diameter of 20-in (~51 cm). This article describes the story of the Kamiokande experiment from the preparation stage to the observation of the first candidate events for proton decays.

## 1. Idea of Kamiokande

The GUT idea of strong, weak and electromagnetic interactions emerged in the 1970s [1,2]. These theories predicted nucleon decays with a typical lifetime of $10^{30}$ years, with about 2 orders of magnitude uncertainty in this prediction. If the nucleon lifetime is less than $10^{32}$ years, it should be possible to observe proton and bound neutron decays by a massive detector with a mass of the order of 1 kt.

A workshop on 'The Unified Theory and the Baryon Number in the Universe' was held in KEK in 1979. H. Sugawara, who was one of the organizers of this workshop, made a phone call in late 1978 to M. Koshiba, asking for a talk on a possible proton-decay experiment. M. Koshiba proposed the initial concept of Kamiokande, which was a water Cherenkov detector. A charged particle propagating in a medium with a speed exceeding the speed of light in the medium emits Cherenkov radiation (light). The speed of light in a medium is $(c/n)$, where $c$ is the speed of light in the vacuum and $n$ is the refractive index of the medium. The refractive index for water is approximately 1.34. Cherenkov light is radiated only in a direction defined by $\cos\theta = 1/(n\beta)$, where $\theta$ is the angle at which Cherenkov photons are emitted relative to the direction of the particle motion, and $\beta$ is the velocity of the particle relative to the light velocity in the vacuum. In water, a highly relativistic particle with a velocity very close to $c$ (namely, $\beta$ very close to 1) thus emits Cherenkov light in the direction of about 42 degrees from the direction of the particle motion. The Cherenkov light can be detected by photon counters, such as photomultiplier tubes (PMTs).

The proposed detector is shown in Fig. 1. Koshiba's idea was presented by a member of his research group (Y. Watanabe) at the workshop [3]. It comprised water Cherenkov counters A and B and was intended to be located deep underground. Counter B was planned to be used to detect nucleon decays that occurred in water within the detector. The proposed size was $20 \times 20 \times 5$ m$^3$, containing 2000-t of water. The counter surface was to be covered with 7500 5-in diameter PMTs, one every $40 \times 40$ cm$^2$. The inside of the counter wall was painted black so as to avoid any reflection of Cherenkov light. Thus the direction and the energy of charged particles could be measured. Counter A, which served as a counter to identify atmospheric muon backgrounds that penetrated through the detector, had a size of $25 \times 25 \times 0.1$ m$^3$, and was placed above counter B. A counter or a detector that is used to reject background events is called a 'veto-counter'.

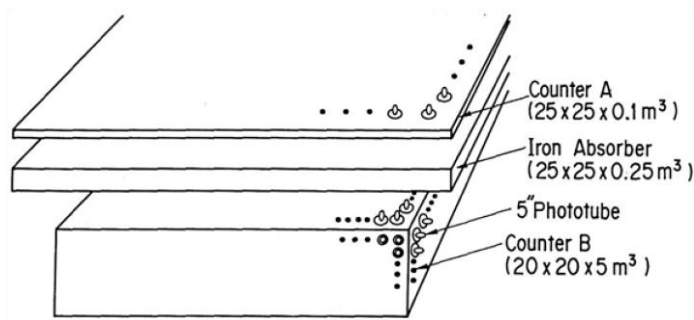

**Fig. 1.** Initial idea of the Kamiokande detector [3].

The inside of this veto-counter was painted white, and the water was dissolved with a wavelength shifter, which absorbs ultraviolet light and re-emits blue light, so as to increase the light yield. A steel plate with a thickness of 25 cm between counters A and B was used to stop any charged particles produced by nucleon decays in counter B.

Figure 2 illustrates how the $p \to \mu^+ + \gamma$ decay would be detected in this detector. Cherenkov light emitted by the muon and an electromagnetic shower initiated by the $\gamma$ is detected by the PMTs instrumented at the inner surfaces of the detector. Each Cherenkov light is recognized as a 'Cherenkov ring'.

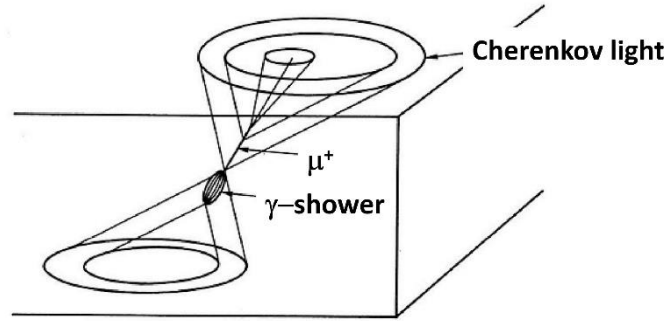

**Fig. 2.** Image of the proton decay into $\mu^+ + \gamma$ in the proposed detector [3].

However, this detector required a huge amount of funding, since it would have required nearly 10 000 PMTs. Hence M. Koshiba thought of making a cubic detector containing 3000 t of water and surrounded by 1000 5-in diameter PMTs. The proposed detector was approved and partially funded by the Ministry of Education. The news then came from the USA that an 8000-t water Cherenkov detector surrounded by 2000 5-in diameter PMTs had been proposed. This was obviously much better than the proposed detector by M. Koshiba as far as the sensitive volume for proton decay was concerned. At this stage the Ministry of Education would not approve any increase of funding. How could this difficulty be resolved within the approved budget? Since the number of PMTs could not be increased, the only possible way out was to increase the sensitivity of each PMT by increasing the photocathode area.

In 1979, M. Koshiba had a discussion with T. Hiruma, who was the president of Hamamatsu TV Co., Ltd. (its present name is Hamamatsu Photonics K. K.), which was one of the companies that was able to produce high-quality PMTs. At the end of the discussion, T. Hiruma agreed to develop large PMTs jointly with the University of Tokyo group.

## 2. Kamiokande site

In the early stage of designing of the 3000-t water Cherenkov detector equipped with 1000 newly developing 20-in PMTs, a cubic detector was considered, being entirely surrounded by a 1 m depth of veto-counter. However, in terms of the stability of the underground cavity, a cylindrically shaped cavity was preferred. Consequently, the detector shape was changed to cylindrical without a veto-counter.

Another important issue was to select the underground site. In order to reduce the cosmic ray background, the detector needed to be located deep underground. From consideration of the cosmic ray muon rate, a depth of approximately 1 km from the surface was required. In addition, the rock condition had to be good enough to excavate a large cavity. Other factors to be considered were the availability of clean water to be used as the primary water for the detector, easy access, the understanding of the mine company operating the underground site.

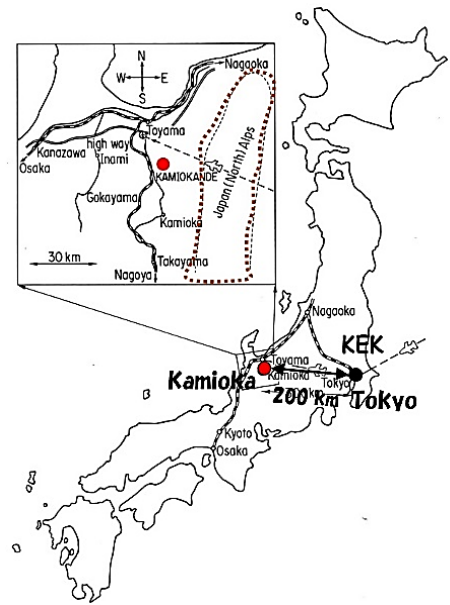

**Fig. 3.** Location of Kamioka

In the 1960s, M. Koshiba, together with T. Suda and Y. Totsuka, had carried out an underground cosmic ray experiment at the Kamioka mine; therefore, the Kamioka mine was one of the candidate sites. After investigating other several candidate sites, the experimental site was finally determined to be at the Kamioka mine located in Central Japan, 220 km north-west of Tokyo (Fig. 3). This experiment was named Kamiokande after the Kamioka Nucleon Decay Experiment.

This mine was actively producing zinc and lead. The underground laboratory was located below the top of a mountain at the depth of 1000 m (see Fig. 4). The average rock density was 2.7 g/cm$^3$; thus, the average rock overburden was approximately 2700 meters-water-equivalent.

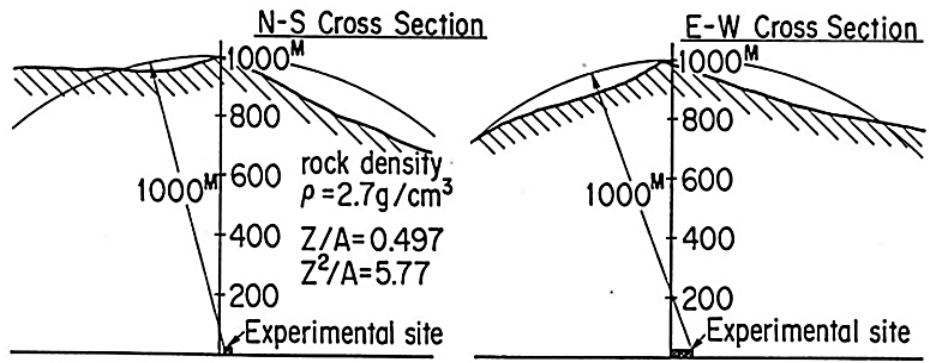

**Fig. 4.** Vertical cross sections of the overlying rocks through the experimental site.

This site had many desirable features. The rock was very solid and stable, and showed no potential hazards with water or air. A well-maintained horizontal access tunnel with a rail track led to the site. A stream of clean natural water of about 2-t/min with 9°C and with pH = 7.8 was available along the tunnel. The temperature in the tunnel was about 13°C all the year around, while the relative humidity was approximately 93%.

Cosmic ray muons with energy higher than ~1.2 TeV can penetrate to this depth. The muon rate through the detector was estimated to be ~0.12/s, where the zenith angle distribution of muons as well as the energy spectrum, the geometry of the mountain surface, and the configuration of the detector were considered.

## 3. Development of 20-in diameter PMT

For a large underground Cherenkov detector for proton-decay studies, it was recognized that the PMT for Cherenkov light detection should have the capabilities of wide photosensitive area, single photoelectron detection, fast timing property, tolerance against high pressure, and long-term stability. The diameter of the PMT was determined to be 20-in by the physics requirement. 20-in PMTs were developed by overcoming several problems with the production, such as glass manufacturing and PMT structure, that had not been encountered before (see Fig. 5) [4,5]. A picture and a plane view of the inner structure of the 20-in PMT are shown in Fig. 6.

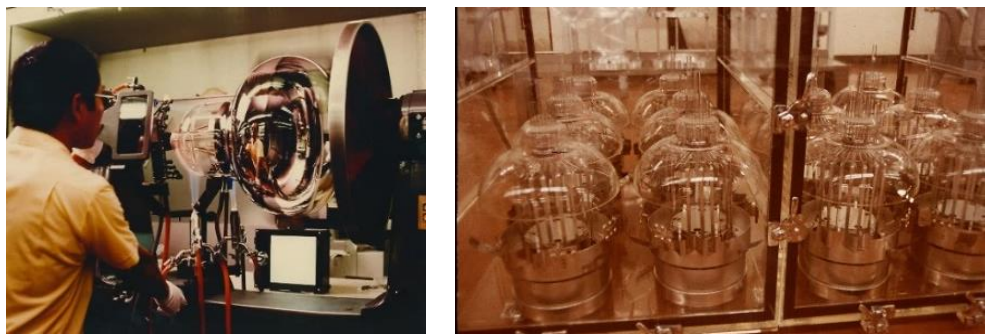

**Fig. 5.** Manufacturing of the 20-in PMT (Courtesy of Hamamatsu Photonics K. K.).

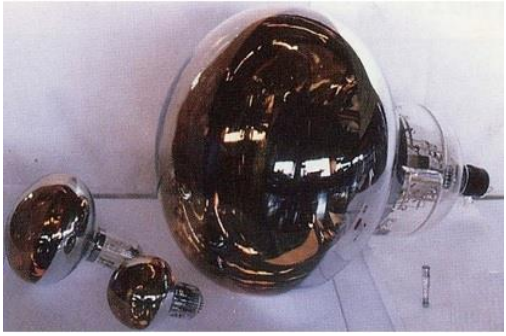 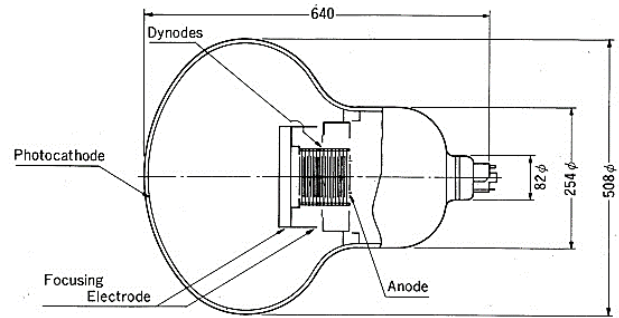

**Fig. 6.** (Left) Picture of the 20-in PMT, compared with 1/2, 5 and 8-in PMTs. (Right) Plane view of the inner structure of the 20-in PMT. (Courtesy of Hamamatsu Photonics K. K.)

The 20-in diameter hemispherical glass window and photocathode provide effective light collection ($2\pi$ solid angle), measurement of timing property, and tolerance against high pressure. The Venetian-blind type dynodes of 13 stages were installed to obtain a large photoelectron collection area as well as high current amplification. The structure and the arrangement of the focusing electrodes, which strongly affect the photoelectron collection efficiency and the timing property, were carefully designed on the basis of electron trajectory analysis. Figure 7 shows the simulation of the electron trajectories from the photocathode to the first dynode with a voltage difference of 800 V. Photoelectrons are traced under an initial energy of 0.5 eV and emission angles of 0º and ±90º.

Several characteristics of the PMT were measured with the 13-stage voltage divider (8-1-1-. . .-1), which provides optimum overall operation.

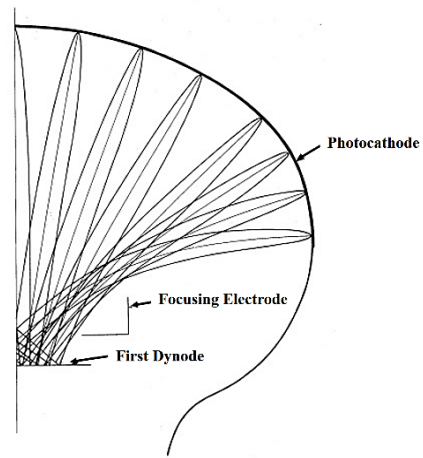

**Fig. 7.** Computer simulation of electron trajectories.

The gain of the PMT defined by the ratio of the anode output current to the emitted photocathode current was to be $10^7$ to detect single photoelectron events. The values of the applied voltage for obtaining a $10^7$ gain were distributed around 2000 V. From this result the PMTs were able to be handled with standard techniques. The dark pulse counting rate should be, of course, low for a better S/N ratio. The discrimination level of 1/4 photoelectron equivalent was set for the elimination of tiny dark pulses. The counting rate was found to lie between 5 kcps (kilo counts per second) and 70 kcps at a temperature of 25 ⁰C. The dark pulse was mainly due to thermal emission from the photocathode, depending on operating temperature. The average value was around 20 kcps at 25 ⁰C. This dark pulse rate is satisfactorily small from the experimental point of view.

When a large PMT is constructed, it is natural to ask whether a uniform sensitivity over the whole photocathode area is obtained or not. This uniformity is called anode uniformity and depends mainly on two factors, i.e. the uniformity of the photocathode quantum efficiency and that of the collection efficiency between the photocathode and the first dynode. The photocathode was illuminated by a light spot through an optical fiber and the PMT was rotated to change the illuminated point on the photocathode around its spherical center. Figure 8 shows the uniformity of the photocathode quantum efficiency (left) and the anode uniformity (right). The notations *X* and *Y* in Fig. 8 represent the rotation direction of the PMT axis. The light position ($\theta$) shows the angle between the incident optical axis and the PMT axis. From Fig. 8 it is found that the change of anode uniformity is within 40%. The PMT has a good uniformity for its large dimension.

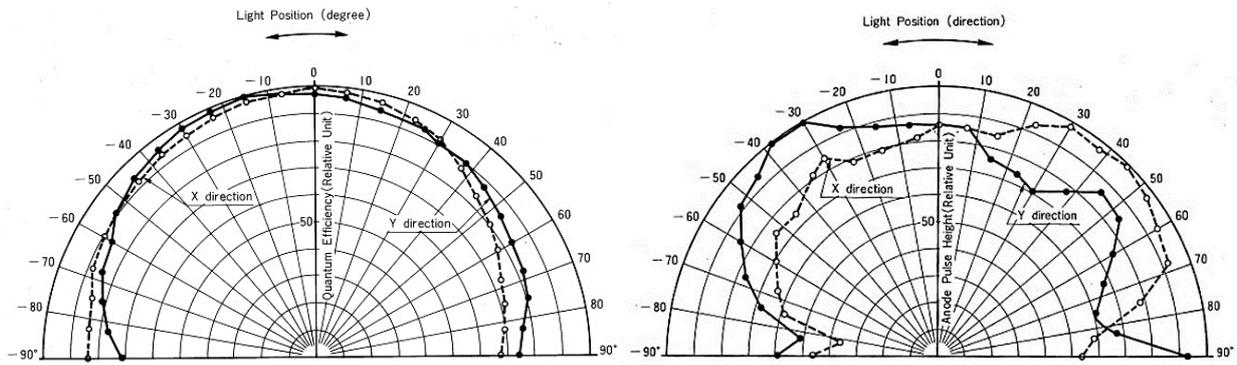

**Fig. 8.** (Left) Photocathode uniformity. (Right) Anode uniformity.

The transit time spread (distribution of transit time for a single PMT) is an important parameter when timing information is required. For a typical 20-in PMT, the transit time spread was found to be 7 ns as the full width at half-maximum (Fig. 9).

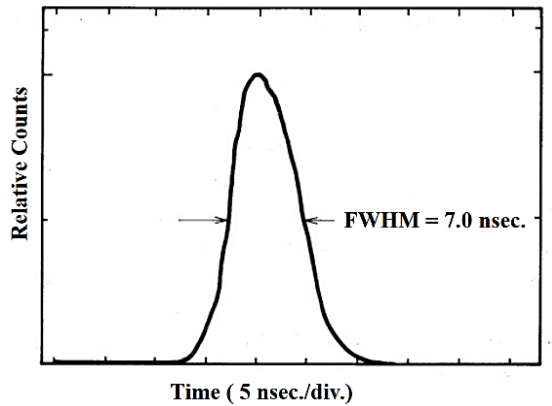

**Fig. 9.** Transit time spread for a single photoelectron pulse.

The 20-in PMT has a weak point; it is very sensitive to magnetic fields. The earth-magnetic field at Kamioka is ~300 mG. This field should be carefully considered because it leads to a systematic error in the total energy measurement. To correct this effect, we adopted two independent precautionary measures. One is to cover each PMT with two special magnetic shields as shown in Fig. 10. The front is a hemispherical mesh of µ-metal (Ni 78 %), which can transmit 88 % of the total incident light. The back is a cone of µ-metal with 0.35-mm thickness. The total shielding factor was found to be 3-8 depending on the relative orientation of the PMT with respect to the earth-magnetic field.

The other is to provide a system of current networks around the whole water tank as shown in Fig. 11. Considering these two corrections, the magnetic field inside the 20-in PMT was less than 30 mG even at the worst position inside the tank. Thus, it was concluded that the gain of each tube does not deviate by more than 3%.

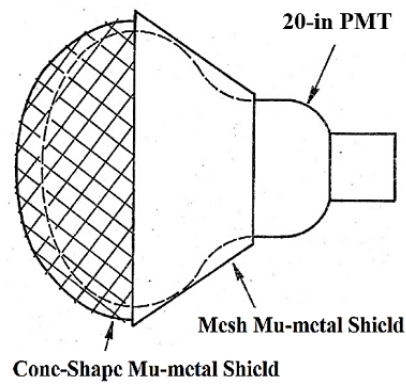

**Fig. 10.** Magnetic shield for the 20-in PMT.

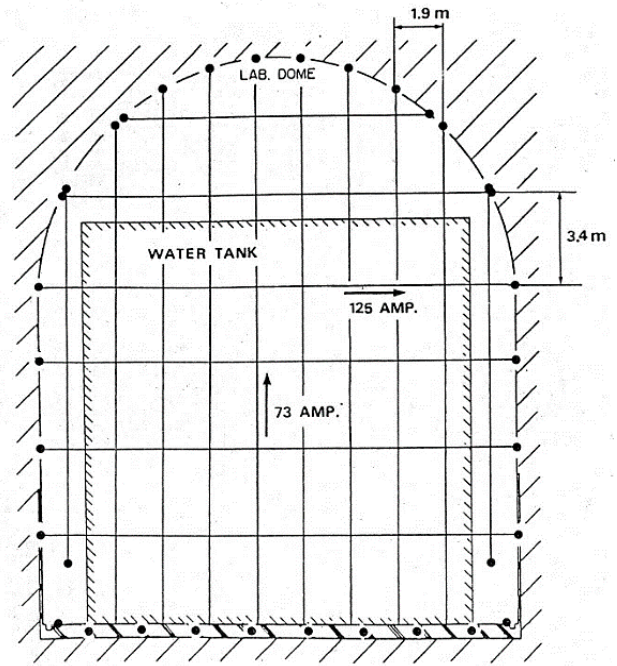

**Fig. 11.** System of current networks to compensate for the earth-magnetic field.

There were no serious problems in using 20-in PMTs as detectors of Cherenkov radiation in Kamiokande. The designed properties were successfully accomplished. The characteristics of the 20-in PMTs are summarized in Table l.

**Table 1.** Characteristics of the 20-in PMT.

| | |
|---|---|
| Photocathode area | 20-in diameter |
| Shape of photocathode | Hemispherical |
| Photocathode material | Bi-alkali |
| Window material | Borosilicate glass Hario-32 (4-5 mm thick) |
| Dynodes | Venetian blind, 13 stages |
| Bleeder chain | 8-1-1⋯-1 |
| Gain | $10^7$ |
| Quantum efficiency | 22% at $\lambda = 400$ nm |
| Dark current | 150 nA at gain $10^7$ |
| Dark pulse count | 15 kHz at gain $10^7$ |
| Mean transit time | 90 ns |
| Transit time spread | 7 ns FWHM (full width at half-maximum) |
| After pulse rate | Less than 1% per photoelectron |

# 4. Design and construction of the Kamiokande detector

4.1 *Experimental site*

A systematic examination of rock condition at the site and a detailed design for engineering to excavate the cavity, as well as for construction of the underground laboratory, were completed by January 1982. Then, the excavation of the large-volume cavity (19 m diameter, 23 m high at the center and 20 m high at the edges) started. It took about 10 months to complete the excavation. A schematic view of the entire experimental area is shown in Fig. 12. The flat spaces in the tunnels near the cavity were used for the data-taking electronics room at the top and for the water-purification system at the bottom.

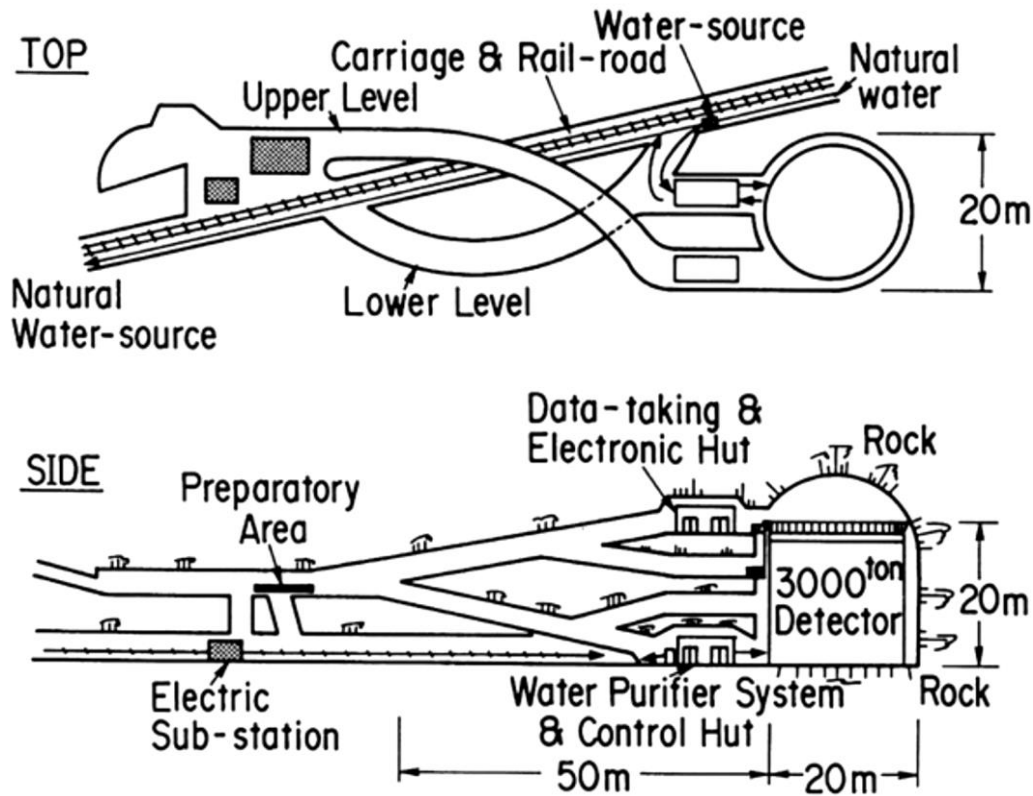

**Fig. 12.** Schematic view of the entire experimental area of Kamiokande.

4.2 *Water tank and PMT support structure*

Construction of the 3000-t water tank followed immediately after the cavity excavation, and continued until March 1983. The pure-water tank consisted of a cylindrical steel tank of 15.5 m in diameter and 16 m in height. The steel was 12 mm thick at the bottom. The thickness was gradually reduced toward the top, ending in a 4.5 mm thickness at the top. The inner walls of the tank were painted with a specially prepared black-colored epoxy. The thickness of epoxy was more than 250 $\mu$m to maintain water purity. Figure 13 depicts the Kamiokande detector.

The Cherenkov light was detected by a total of 1000 20-in PMTs installed on the inner walls of the 3000-t water tank. These PMTs were uniformly distributed with a density of 1 PMT/m$^2$. PMTs were individually supported by a frame in the form of a matrix array bolted onto the wall of the steel tank. With this PMT arrangement, 20 % of the inner wall of the tank was covered by photosensitive cathode. This offered a good energy resolution, approximately 4% for a l GeV electron shower event. Consequently, good background rejection and good identification of various modes of nucleon decays were expected. Figure 14 shows a typical example of a simulated Cherenkov-ring pattern appearing on the inner walls of the tank for a proton decay into e$^+\pi^0$ (left) and the response of the PMT array for this event (right).

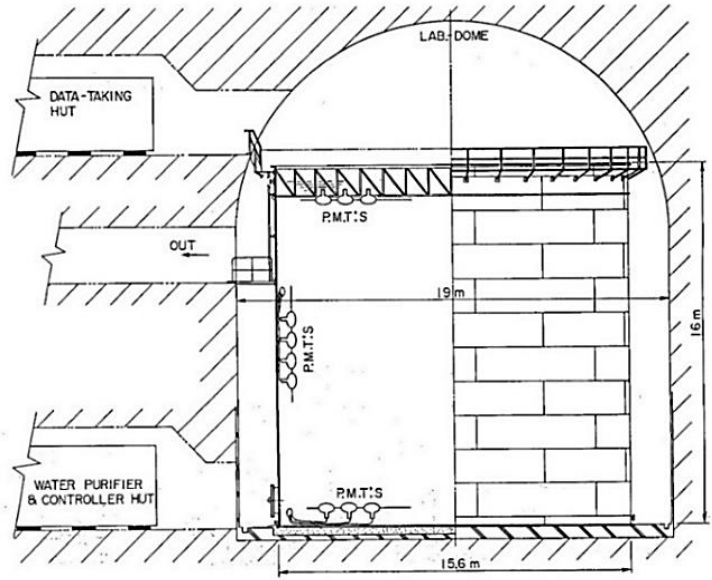

**Fig. 13.** Original Kamiokande detector.

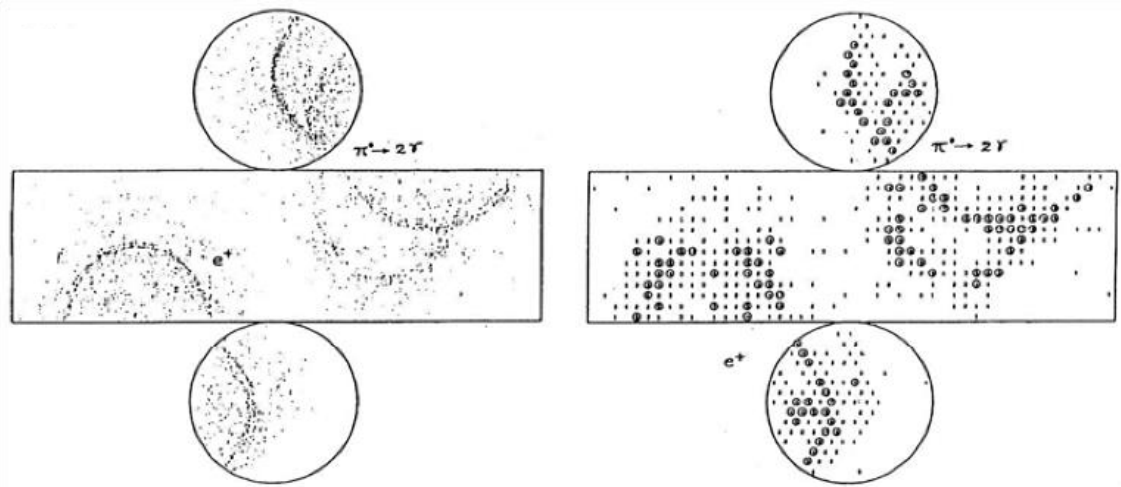

**Fig. 14.** (Left) Simulated Cherenkov light pattern and (right) Kamiokande response.
PMTs that observed more than four photoelectrons are marked by circles.

4.3 *Water supply and purification system*

We investigated the natural water flowing in the ditch at the site and found that it was clean enough for a primary source of water for the Cherenkov detector. The levels of undesirable chemical contents such as $Fe^{++}$, $Ni^{++}$, and $Co^{++}$ ions were found to be too low to be of any effect at less than 0.02 ppm. Other contaminating chemicals or organic materials such as bacteria were also much less than those of usual city drinking water. The temperature of the rock cavity at the mine was about 12°C and was almost constant all the year around. This relatively low temperature was very helpful to suppress undesirable bacteria multiplication in the water inside the tank. Having this excellent primary water supply and environment, the design of the water-purification system was considerably simplified. In the first stage of the experiment, the system consisted of pre-filtration (25 μm) + deionization + fine filtration (0.22 μm). By this system water with a transparency of > 25 m at λ ~ 470 nm was obtained. If necessary, reverse osmosis was planned to be added in the later stage. The maximum processing rate was 10 m$^3$/hr, and it took half a month to fill up the 3000-t tank. Water was continuously circulated and purified during normal operation. The purity of the water was continuously monitored by measuring resistivity and density of microparticles. The transparency of the water was also checked using laser light. Figure 15 shows the water flow scheme in the water-purification system.

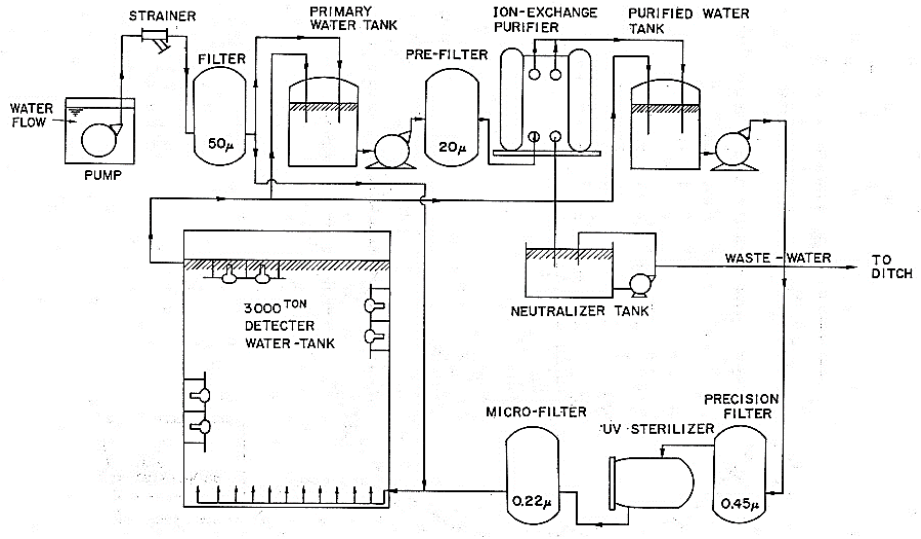

**Fig. 15.** Water-purification system

4.4 *Electronics data acquisition and calibration system*

The principle parts of the electronics and data-taking system were linear adders, ADCs, transient digitizers, and a PDP-11/60 on-line computer. The signal from each PMT first went to a linear adder and then split into two. One went to a 12-channel summing amplifier and the other to the ADC system, which digitized the total charge from each PMT with 15-bit resolution. After three summing-amplifier steps, all the currents from 1000 PMTs were summed up. The total sum of the PMT currents provided the energy trigger. An event was triggered when the total energy deposition exceeded 100 MeV (~130 photoelectrons equivalent). The threshold could be reduced to ~ 20 MeV.

To identify a muon by observing a delayed electron from $\mu \rightarrow e\nu\bar{\nu}$ decays, the signals were summed up and digitized over a time of ~5 μs by sampling every 10 ns with a pulse height full width of 9 bits. Figure 16 shows a block diagram of the electronics and data acquisition system in Kamiokande.

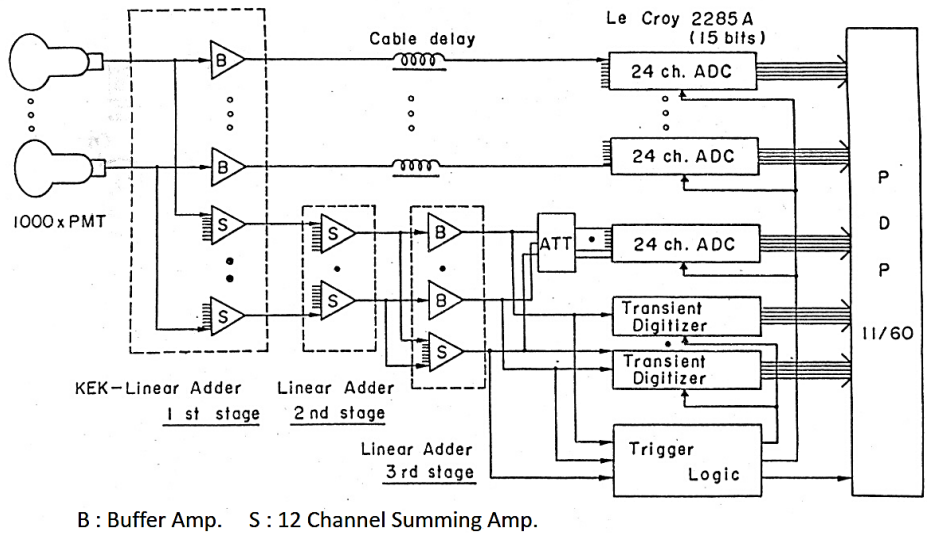

**Fig. 16.** Block diagram of the electronics and data acquisition system.

The gain of the 1000 PMTs in the tank was monitored by a flashing scintillating ball hanging at the center of the tank and at four different spots when necessary. This plastic scintillating ball contained 50 ppm wavelength-shifter materials and emitted isotropic light with a spectrum of λ = 390~500 nm when excited by a light of λ = 300~410 nm generated by a Xe lamp with a proper filter and a fiber light guide. This scintillating ball monitor system was calibrated independently by the standard PMT.

The construction of the water-purification system and the data- acquisition electronics system began in early 1983. All of the necessary instruments were delivered to the mine by the end of March, 1983.

4.5 *Installation of 1000 20-in PMTs*

During detector construction, 1000 20-in PMTs were delivered to the mine from various storage places. The installation of 1000 PMTs into the tank began in April l983. It took approximately three months to install the PMTs into the tank. At first, the PMTs in the bottom plane and the lowest two side rows were mounted onto the support frame by hand. Then, water was poured into the tank, and the installation of the side-wall PMTs was carried out on floating rubber boats (see Fig.17). The water was leveled up after completing all 48 tubes in one row on the side wall. It took two working days to install 48 PMTs. Finally, the PMTs at the top plane were installed using a crane.

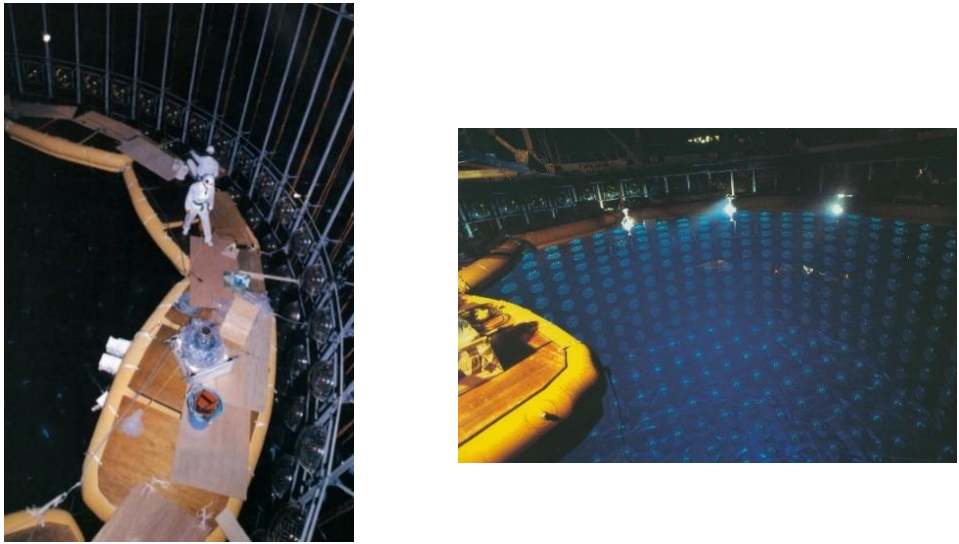

**Fig. 17.** PMT installation on the side wall of the tank.

The detector was ready in early July 1983. The inner view and a sketch of Kamiokande detector are shown in Fig. 18. The design and construction work of Kamiokande were presented in several conferences [6 - 9].

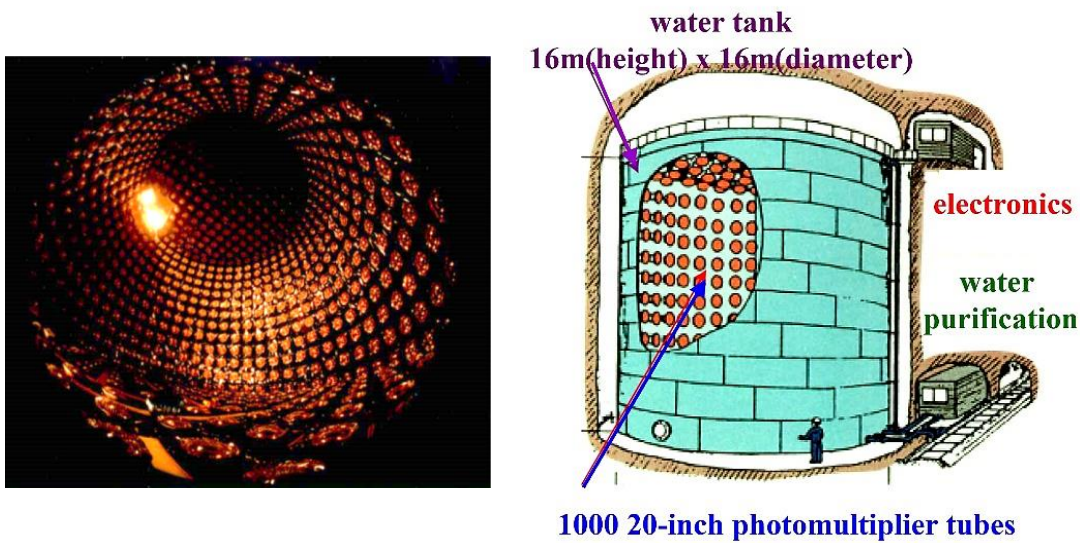

**Fig. 18.** Kamiokande 3000-t water Cherenkov detector: (left) inner view and (right) sketch.

# 5. Launch of the Kamiokande data-taking

The Kamiokande data-taking started on 6 July 1983. In July and August, much time was spent examining the detector performance and to develop analysis programs for events such as nucleon decays and atmospheric neutrinos. Data tapes were regularly sent from Kamioka to the University of Tokyo. The computer system at ICEPP (International Center for Elementary Particle Physics) at the University of Tokyo was fully utilized for the data analysis.

Data accumulated over 176 days from 6 July 1983 were analyzed to reveal the first result. After subtracting the suspension of data recording due to adjustment of experimental equipment, power outages, computer data collection time, etc., the data from 134.6 days were selected as available measurements. During this period, $5.8\times10^6$ events were recorded, about 80% of which were cosmic ray muons produced in the atmosphere, penetrating deep underground, and passing through the detector. After selecting events with both a total energy below 1300 MeV and with all particles contained inside the detector, 65 726 events remained.

Then, displaying events on a graphic terminal, scanning judgment by the human eye was performed to choose events that occurred internally more than 2 m from the PMT surface. As a result, 57 events eventually remained as candidates for atmospheric neutrino reactions and proton decays.

To identify proton-decay events from among these 57 events, kinematic conditions for the particles produced and the event were applied. Consequently, two events were found as candidates for proton decays. One was the proton decay of $p \rightarrow \mu^+\eta$ ($\eta \rightarrow 2\gamma$). The other was $p \rightarrow e^+\omega$ ($\omega \rightarrow \pi^+\pi^-\pi^0$).

## 5.1 *Proton decays?*

The Cherenkov light and ring pattern of the candidate event of $p \rightarrow \mu^+\eta$ ($\eta \rightarrow 2\gamma$) are shown in Fig. 19. In this figure the amount of light detected by each PMT is illustrated by the size of the circle on the expanded plane of the cylindrical Kamiokande detector. ② in this figure was identified with $\mu^+$, and ① and ③ were considered as $2\gamma$ decayed from η. However, there were difficulties such as the slightly large deviation from the momentum conservation of $\mu^+$ and η, and also the ambiguous discrimination of $\mu^+$ and $\pi^-$.

The candidate event of $p \rightarrow e^+\omega$ ($\omega \rightarrow \pi^+\pi^-\pi^0(\gamma\gamma)$) is shown in Fig. 20. Since it is not possible to uniquely determine which ring is which particle, it was analyzed with a combination of all possibilities. The frequency with which atmospheric neutrinos create such events was calculated as 0.2 events/134.6 days.

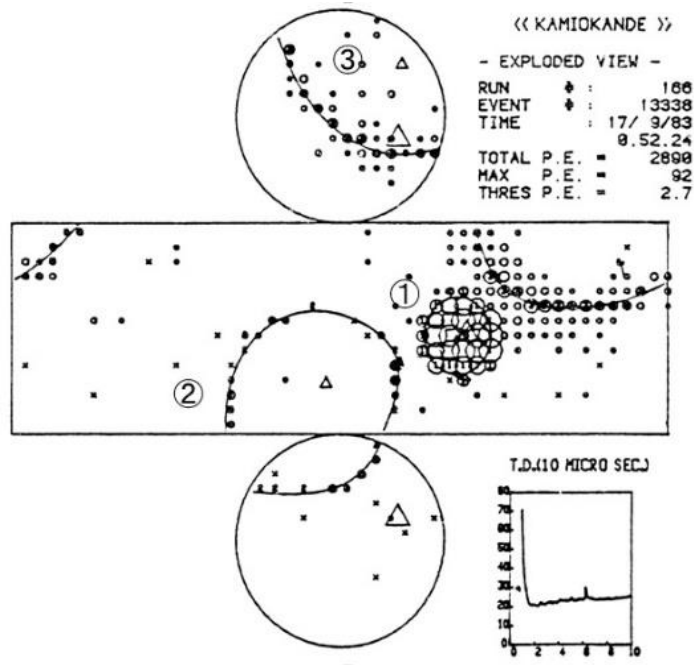

**Fig. 19.** The candidate event of $p \rightarrow \mu^+\eta$ ($\eta \rightarrow 2\gamma$).

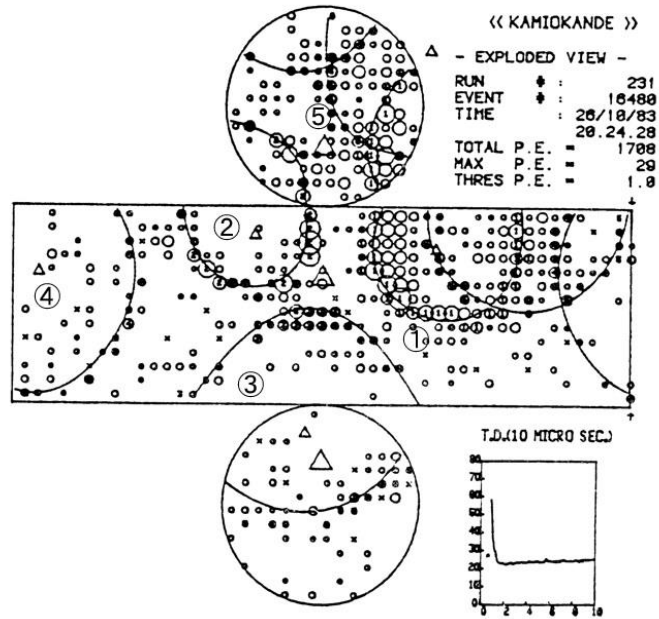

**Fig. 20.** The candidate event of $p \rightarrow e^+\omega$ ($\omega \rightarrow \pi^+\pi^-\pi^0(\gamma\gamma)$).

From the above results, the shortest decay mode of $p \rightarrow e^+\pi^0$ predicted by SU(5) was not detected. Although the discrepancy between such a theory and experiment is a mystery, considering the two candidate events from the 57 events contained inside the detector, the remaining events contained inside the detector originated from atmospheric neutrinos, i.e. background events of proton-decays. The signal-to-noise ratio was 1 : 30. This required an improvement in the ability to identify atmospheric neutrino events and to increase the data-taking time. This was the conclusion from the first result from Kamiokande. This result was reported by M. Koshiba at the International Conference on High Energy Physics in 1984 [10].

## 6. Summary

Although there was no solid evidence of proton decays, the performance of Kamiokande was closely evaluated in conferences, especially the detailed identification methods of atmospheric neutrinos and proton decays surpassing other proton-decay experiments. Soon after the International Conference on High Energy Physics in 1984, M. Koshiba started to upgrade Kamiokande, making it possible to detect lower-energy events, in particular for solar neutrinos [11]. M. Koshiba also proposed a 32 000-t large-water Cherenkov detector, JACK (Japan-America Collaboration at Kamioka) at the end of 1983 [12].

Kamiokande was extended to mean not only the Kamioka Nucleon Decay Experiment, but also the Kamioka Neutrino Detection Experiment which led to both neutrino astronomy and finite neutrino masses. The whole story of the activities at Kamiokande is summarized in Refs. [13,14].

Finally, let us introduce one of M. Koshiba's didactic messages. 'Experimental equipment should incorporate as many distinctive devices as possible. That way, even if you fail to catch the prey you are aiming for, you will have a chance to catch other prey.' This was precisely the case for the 20-in PMT at Kamiokande.

**Acknowledgements**
The author would like to acknowledge the colleagues in Kamiokande and also the generous cooperation of the Kamioka Mining and Smelting Co. and the Hamamatsu Photonics K. K.

# References

[1] J.C. Pati and A. Salam, Phys. Rev. D **10**, 275 (1974).

[2] H. Georgi and S.L. Glashow, Phys. Rev. Lett. **32**, 438 (1974).

[3] Y. Watanabe, Proc. Workshop Unified Theory and the Baryon Number in the Universe, KEK (High Energy Physics Laboratory in Japan) Report, KEK-79-18 (1979), p. 53.

[4] T. Hayashi, Proc. 1981 INS (Institute of Nuclear Study, University of Tokyo) Int. Symp. Nuclear Radiation Detectors (Tokyo, Japan, 1981).

[5] H. Kume, S. Sawaki, M. Ito, K. Arisaka, T. Kajita, A. Nishimura and A. Suzuki, Nucl. Instrum. Meth. **205**, 443 (1983).

[6] T. Suda et al., Proc. Neutrino '81, Hawaii, Maui, (1981), Vol. **1**, p. 224.

[7] H. Ikeda et al., Proc. Third Workshop Grand Unification (University of North Carolina, Chapel Hill, N. C., 1982).

[8] M. Koshiba, Proc. Of the 21st Int. Conf. High Energy Physics (Paris, 1982).

[9] M. Koshiba, Proc. Topical Symp. High Energy Physics (University of Tokyo, Tokyo, 1982), p. 245.

[10] M. Koshiba et al., Proc. 22nd Int. Conf. High Energy Physics (Leipzig, East Germany, 1984), Vol. II, p. 67.

[11] M. Koshiba et al., Proc. International Colloquium on Baryon Nonconservation (ICOBAN '84) (Park City, Utah, 1984).

[12] M. Koshiba, Proc. of Workshop on Grand Unified Theories and Cosmology, KEK Report, KEK-84-12 (1983), p. 24.

[13] M. Fukugita and A. Suzuki, Physics and Astrophysics of Neutrinos (Springer, Berlin, 1994).

[14] T. Kajita, M. Koshiba, and A. Suzuki, Eur. Phys. J. H **37**, 33 (2012).